\documentclass[%
reprint,
preprintnumbers,
showkeys,
amsmath,amssymb,
aps,
prl,
floatfix,superscriptaddress,a4paper,left=1.5cm,right+1.5cm,
]{revtex4-1}

\usepackage[english]{babel}
\usepackage[utf8]{inputenc}
\usepackage{graphicx}
\usepackage{hyperref}
\usepackage{xcolor}
\usepackage{amsmath}
\usepackage{amsfonts}
\usepackage{amssymb}
\usepackage{dsfont}
\usepackage{siunitx}
\usepackage{xcolor,import}
\usepackage{transparent}
\usepackage{sidecap} % For side caption
\usepackage{newtxtext,newtxmath}
\usepackage[T1]{fontenc}

\hypersetup{
    pdftoolbar=true,               % show Acrobat’s toolbar?
    pdfmenubar=true,               % show Acrobat’s menu?
    pdffitwindow=false,            % window fit to page when opened
    pdfstartview={FitH},           % fits the width of the page to the window
    pdftitle={},                   % title
    pdfnewwindow=true,             % links in new window
    colorlinks=true,               
}

\begin{document}

\title{Controlled multi-photon subtraction with cascaded Rydberg superatoms as single-photon absorbers}

\author{Nina Stiesdal}
\affiliation{Department of Physics, Chemistry and Pharmacy, Physics@SDU, University of Southern Denmark, 5230 Odense M, Denmark}
\author{Hannes Busche}
\affiliation{Department of Physics, Chemistry and Pharmacy, Physics@SDU, University of Southern Denmark, 5230 Odense M, Denmark}
\author{Kevin Kleinbeck}
\affiliation{Institute for Theoretical Physics III and Center for Integrated Quantum Science and Technology,\\ University of Stuttgart, 70550 Stuttgart, Germany}
\author{Jan Kumlin}
\affiliation{Institute for Theoretical Physics III and Center for Integrated Quantum Science and Technology,\\ University of Stuttgart, 70550 Stuttgart, Germany}
\author{Mikkel G. Hansen}
\affiliation{Department of Physics, Chemistry and Pharmacy, Physics@SDU, University of Southern Denmark, 5230 Odense M, Denmark}
\author{Hans Peter B\"uchler}
\affiliation{Institute for Theoretical Physics III and Center for Integrated Quantum Science and Technology,\\ University of Stuttgart, 70550 Stuttgart, Germany}
\author{Sebastian Hofferberth}
\email{hofferberth@sdu.dk}
\affiliation{Department of Physics, Chemistry and Pharmacy, Physics@SDU, University of Southern Denmark, 5230 Odense M, Denmark}
\affiliation{Institut für Angewandte Physik, University of Bonn, 53115 Bonn, Germany}

\date{\today}

\begin{abstract}
The preparation of light pulses with well-defined quantum properties requires precise control at the individual photon level.
Here, we demonstrate exact and controlled multi-photon subtraction from incoming light pulses. 
We employ a cascaded system of tightly confined cold atom ensembles with strong, collectively enhanced coupling of photons to Rydberg states.
The excitation blockade resulting from interactions between Rydberg atoms limits photon absorption to one per ensemble and engineered dephasing of the collective excitation suppresses stimulated re-emission of the photon.
We experimentally demonstrate subtraction with up to three absorbers.
Furthermore, we present a thorough theoretical analysis of our scheme where we identify weak Raman decay of the long-lived Rydberg state as the main source of infidelity in the subtracted photon number.
We show that our scheme should scale well to higher absorber numbers if the Raman decay can be further suppressed.
\end{abstract}

\maketitle

%=================%
%  Introduction
%=================%

\noindent Future optical quantum technology relies on precise control over the quantum state of light.
Deterministic removal of exactly one, or more generally exactly $n_{\mathrm{sub}}$, photons enables applications such as state-preparation for optical quantum simulation and computing \cite{Milburn1989,Fiurasek2005,Neergaard2006,Takahashi2010,Chang2008} or quantum-enhanced metrology \cite{Braun2014}.
Photon subtraction can also give insight into more fundamental aspects of quantum optics \cite{Parigi2007,Kumar2013}.
Heralded single photon subtraction \cite{Calsamiglia2001} can be implemented using highly imbalanced beamsplitters \cite{Parigi2007,Ourjoumtsev2006}, but the probabilistic nature limits the scalability of this approach \cite{Neergaard2006,Ourjoumtsev2006,Zavatta2008}.
Individual absorbers like a single two-level atom in free space seem well-suited for photon subtraction, as they are saturated by just one photon, but this approach is limited by weak atom-photon coupling, stimulated emission, and short lifetimes of the saturated state.
These problems can be mitigated by enhancing the atom-light coupling using a resonator, and transfer of the absorber to a third, dark state \cite{Pinotsi2008,Hoi2013,Gea-Banacloche2013} not coupled to the incoming light as demonstrated with single atoms coupled to a microsphere resonantor \cite{Rosenblum2016}.

Strong photon-emitter coupling can also be achieved without optical resonantors in atomic ensembles with collectively excited and long-lived Rydberg states, also referred to as Rydberg superatoms \cite{ParisMandoki2017}.
Rydberg atoms interact strongly with each other \cite{Saffman2010} and the resulting excitation blockade \cite{Lukin2001} can be mapped onto light fields to create strong optical non-linearites at the single-photon level \cite{Pritchard2010,Dudin2012,Peyronel2012,Firstenberg2016,Murray2016}.
This has enabled many technical applications such as single-photon sources \cite{Ripka2018,Ornelas-Huerta2020}, optical transistors \cite{Gorniaczyk2014,Tiarks2014}, removal of photons from stored light pulses \cite{Gorshkov2013,Murray2018}, and photon-photon quantum gates \cite{Tiarks2019} with recent efforts to combine these into multi-node networks \cite{Thompson2016,Busche2017,Distante2017}. 
Photon subtraction can also be realised using Rydberg superatoms as saturable single photon absorbers \cite{Honer2011,Tresp2016} combining the blockade, which prevents multi-photon absorption, with rapid dephasing of the superatom into dark collective states to avoid stimulated re-emission. 

In this work, we demonstrate a cascaded quantum system of up to three Rydberg superatom absorbers for controlled subtraction of specific photon numbers from a light pulse.
Importantly, our free-space approach avoids the intrinsic insertion loss of optical fibres or waveguides.
In addition to demonstrating controlled multi-photon subtraction, we analyse the performance of our system and find that Raman decay is the main source of deviations from the ideal absorber behaviour.
This is supported by a detailed theoretical analysis, which also shows that scaling beyond $n_{\mathrm{sub}}=3$ absorbers with high probability to subtract exactly $n_{\mathrm{sub}}$ photons is realistic as long as the decay can be further suppressed and a sufficiently high single-photon coupling is maintained.

%=================%
%  Principle and Setup
%=================%

\begin{figure}
    \centering
    \includegraphics{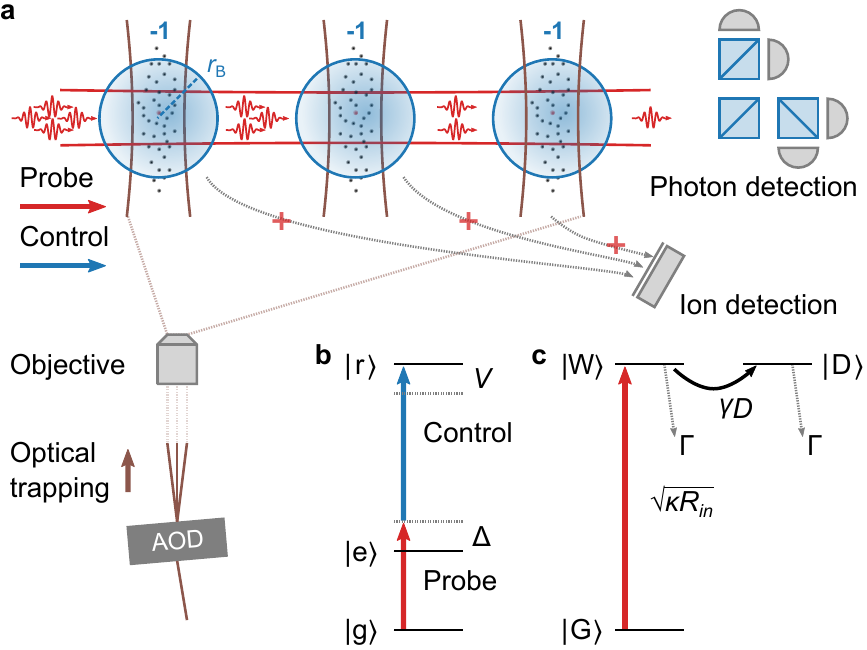}
    \caption{\textbf{Realisation of up to three cascaded single-photon absorbers using Rydberg superatoms. a} To create $n_{\mathrm{sub}}$ saturable superatom absorbers, we place $n_{\mathrm{sub}}$ ensembles of cold $^{87}$Rb atoms in the path of a tightly focussed probe beam. Using an acousto-optical deflector (AOD), we can control the number and position of the optical traps that tightly confine the ensembles below the Rydberg blockade radius $r_{\mathrm{B}}$ along the probe direction. \textbf{b} Within $r_{\mathrm{B}}$ strong van-der-Waals interactions $V$ restrict each ensemble to a single Rydberg excitation as the probe photons and a control field couple $\lvert g \rangle = \lvert 5S_{1/2},F=2,m_F=2 \rangle$ to a Rydberg state $\lvert r\rangle = \lvert 121 S_{1/2},m_J=1/2\rangle$ via $\lvert e \rangle = \lvert 5P_{3/2},F=3,m_F=3 \rangle$ in a Raman scheme with detuning $\Delta/2\pi\approx 100\,\mathrm{MHz}$ and thus to the absorption of a single photon at a time. The transmitted probe pulses are coupled into a single-mode optical fibre (not shown) and detected on four single-photon counters in a Hanbury-Brown Twiss configuration.
    \textbf{c} Representation of the absorber as an effective three level system in terms of singly excited collective states following adiabatic elimination of $\vert e\rangle$. Strong dephasing $\gamma_D$ from the bright excited state $\vert W\rangle$, with strong coupling $\sqrt{\kappa R_{\mathrm{in}}}$ from the ground state $\vert G\rangle$, into dark excited states $\vert D\rangle$ prevents stimulated re-emission of the absorbed photon and the absorption of further photons until it is subject to Raman decay $\Gamma\ll\gamma_D,\kappa$.}
    \label{fig:figure1}
\end{figure}

\noindent\textbf{Implementation.}
Figure 1a illustrates the implementation of individual Rydberg superatoms as a saturable single photon absorber \cite{Honer2011,Tresp2016}. 
Weak pulses of probe light at $\lambda_p\approx 780\,\mathrm{nm}$, from which photons are to be subtracted, propagate through a small, optically thick ensemble of optically trapped $^{87}$Rb atoms.
In combination with a strong, co-propagating control field at $\lambda_c\approx 480\,\mathrm{nm}$, the probe light couples the atomic ground state $\lvert g \rangle = \lvert 5S_{1/2},F=2,m_F=2 \rangle$ to a Rydberg state $\lvert r\rangle = \lvert 121 S_{1/2},m_J=1/2\rangle$ via $\lvert e \rangle = \lvert 5P_{3/2},F=3,m_F=3 \rangle$ in a Raman scheme (Fig. 1b).
As a result of the Raman detuning $\Delta/2\pi \approx 100\,\mathrm{MHz}$, probe photons are only absorbed by the ensemble if the control field is tuned onto Raman resonance.
Strong van-der-Waals interactions between Rydberg atoms lead to the blockade effect that suppresses multiple Rydberg excitations for atoms separated by $r<r_{\mathrm{B}}$ where $r_{\mathrm{B}}$ is the blockade radius that characterises the volume inside which the energy shift $V=C_6/r^6$ defined by the van-der-Waals coefficient $C_6$ exceeds the excitation linewidth.

If the radial probe size ($1/e^2$-waist radius $\approx 6.5\,\mathrm{\mu m}$) and the axial extent of the ensemble constrain the excitation volume to a single Rydberg excitation, the superatom is saturated after absorbing just one photon. 
Consequently, all $N$ atoms in the excitation volume share the excitation in a collective bright state $\lvert W \rangle =  \sum_{j}\lvert g_1 g_2... r_j...g_N\rangle/\sqrt{N}$ with strongly enhanced collective coupling $\sqrt{\kappa R_{\mathrm{in}}}=\sqrt{R_{\mathrm{in}} N}g_0\Omega_c/(2\Delta)$ from the collective ground state $\lvert G \rangle = \lvert g_1 g_2...g_j... g_N\rangle$ \cite{Dudin2012b,Ebert2014,Weber2015,Zeiher2015,ParisMandoki2017}. 
Here, $g_0$ is the single-photon-single-atom coupling strength, $R_{\mathrm{in}}$ the incoming probe photon rate, and $\Omega_c$ the control Rabi frequency.
Following absorption of a photon, $\lvert W \rangle$ dephases with rate $\gamma_D$ into a manifold of $N-1$ collective dark states $\{\lvert D \rangle\}$ that are orthogonal to $\lvert W \rangle$ and no longer couple to the probe such that stimulated emission is suppressed, while maintaining the blockade.
Besides dephasing, the excited collective states are also subject to decay of $\lvert r\rangle$ with Raman decay $\Gamma=\Gamma_e\Omega_c^2/(2\Delta)^2$ as the dominant contribution, with $\Gamma_e$ being the natural linewidth of $\lvert e\rangle$.
Following adiabatic elimination of $\lvert e\rangle$, the superatom dynamics can be described in terms of just $\lvert W\rangle$, $\lvert G \rangle$, and $\lvert D\rangle$, a single dark state into which we condense all collective states in $\{\lvert D \rangle\}$ \cite{ParisMandoki2017,Kumlin2018}. 
This effective three-level system will form the foundation of our theoretical analysis.

To implement multi-photon subtraction, we place $n_{\mathrm{sub}}$ ensembles along the path of the probe and control fields (Fig. 1a) at distances $> 50\,\mathrm{\mu m}\gg r_{\mathrm{B}}$ such that the superatoms do not blockade each other and act independently.
We trap the ensembles at the foci of individual optical trapping beams that intersect perpendicularly with the probe and axially confine each ensemble within $r_{\mathrm{B}}$.
We use an acousto-optical deflector (AOD) to control the position and number of ensembles via the number and frequencies of radio-frequency signals applied \cite{Roberts2014,Barredo2016,Endres2016}.
A reservoir dipole trap (not shown in Fig. 1a) provides additional radial confinement (Methods).
The ensembles have temperatures of $\approx 10\,\mathrm{\mu K}$ with an axial extent $< r_{\mathrm{B}}$ and $N\propto 10^{4}$ depending on $n_{abs}= 3, 2, 1$, respectively.

%=================%
%  3, 2, 1 - subtraction
%=================%
\begin{figure*}
    \centering
    \includegraphics{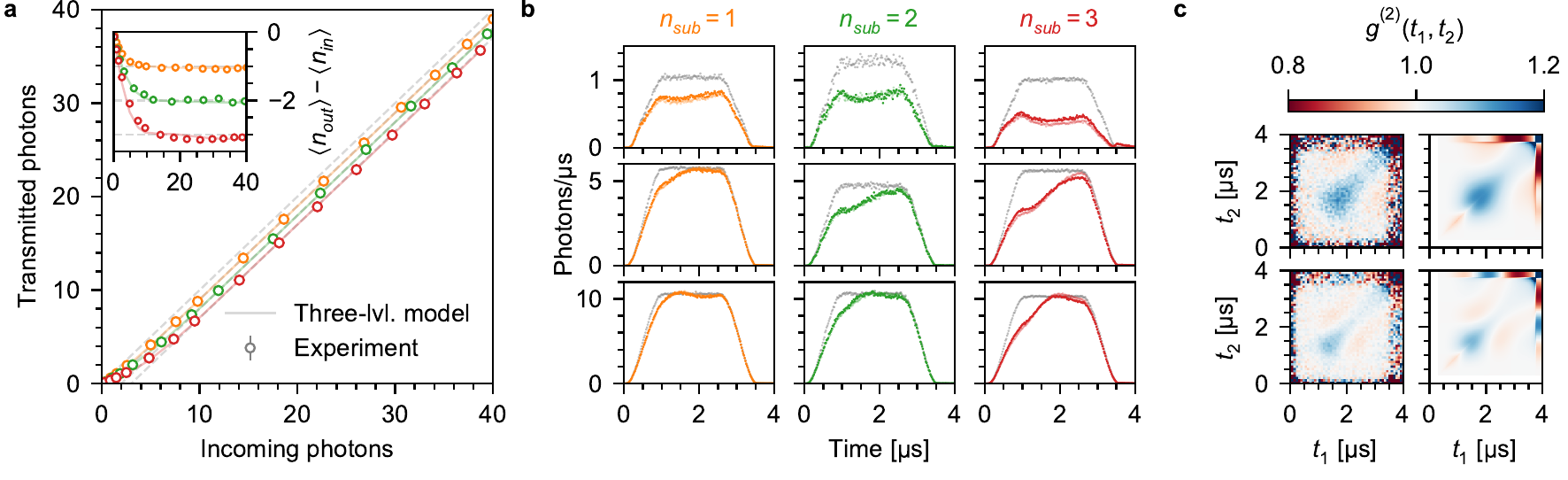}
    \caption{\textbf{Effect of one, two, and three single-photon absorbers on the probe transmission.} \textbf{a} Mean transmitted photons vs. mean incoming photon number. The inset shows the mean subtracted photons for the same data. All data are corrected for off-resonant absorption of the probe light and background counts. Errorbars correspond to one standard error and are smaller than the markers for most datapoints.\textbf{b} Temporal profiles of probe light following transmission through $n_{\mathrm{sub}}=1$ (left), $2$ (center), and $3$ (right) absorbers for different mean photon numbers. The incoming pulses recorded in absence of the superatom absorbers are shown in grey. Besided the experimental data, we also show the results of the three-level model fitted to the experimental data (shaded).
    \textbf{c.} Photon correlations following transmission through three superatom absorbers. Second order correlation functions $g^{(2)}(t_1,t_2)$ for  $R_{\mathrm{in}}\approx 5.5$ (top) and $10\,\mathrm{\mu s}$ (bottom). Besides the experimental data (left), we also show the predictions of the three-level model (right). }
    \label{fig:figure2}
\end{figure*}

\section*{Results}

In the following, we experimentally demonstrate controlled subtraction of up to three photons by placing the corresponding number of absorbers $n_{abs}$ in the probe path.
We measure the transmission for coherent, Tukey shaped probe pulses with a pulse length of $\tau=2.5\,\mathrm{\mu s}$ (FWHM, with $1.0\,\mathrm{\mu s}$ rise/fall time) and mean incoming photon numbers $\langle n_{\mathrm{in}}\rangle\leq 40$ using single-photon counters in a Hanbury-Brown Twiss configuration located behind a single-mode fibre.  
Without control field, we measure combined optical depths of the ensembles of $\approx 11, 16$, and $20$ for $n_{\mathrm{sub}}=1, 2$, and $3$ respectively, and find an only slightly reduced probe transmission of $>0.99$ at finite $\Delta/2\pi \approx 100\,\mathrm{MHz}$, which the data below are corrected for.
As expected, the loss is much lower compared to typical insertion losses into waveguides or optical fibres highlighting a key advantage of the free-space approach.

First, we investigate the difference between $\langle n_{\mathrm{in}}\rangle$ and the mean transmitted photon number $\langle n_{\mathrm{out}}\rangle$ (Fig. 2a).
For $\langle n_{\mathrm{in}}\rangle > 10$, we observe the expected reduction by $n_{\mathrm{sub}}$, while we subtract fewer photons for $\langle n_{\mathrm{in}}\rangle < 10$.
This behaviour is expected, as for low $R_{\mathrm{in}}\propto \langle n_{\mathrm{in}}\rangle$, the pulse area $\sqrt{\kappa R_{\mathrm{in}}}\tau$ does not reach $\pi$ as required to  fully drive the superatom to $\vert W\rangle$ and $\vert D\rangle$.
This becomes particularly evident in the shape of the transmitted pulses for $R_{\mathrm{in}}\approx 1\,\mathrm{\mu s^{-1}}$ (top row in Fig. 2b) with transmission well below one at their end, whereas we observe the onset of saturation for $R_{\mathrm{in}}\approx 5\,\mathrm{\mu s^{-1}}$ (centre row).
Importantly, the duration to reach saturation increases with $n_{\mathrm{sub}}$, because the driving between $\vert G\rangle$ and $\vert W\rangle$ reduces  alongside the probe intensity following each absorber.
For $R_{\mathrm{in}}\approx 10 \,\mathrm{\mu s^{-1}}$ (bottom row), saturation sets in even faster, but we observe a slight oscillation in the subsequent transmission, which reflects the superatom dynamics as the probe drives Rabi oscillations between $\vert G\rangle$ and $\vert W\rangle$ with strong damping due to $\gamma_D$ \cite{ParisMandoki2017}.
To suppress superradiant reemission of absorbed photons in the forward direction after the probe pulse \cite{Moehl2020,Stiesdal2020}, $\gamma_D$ has to be sufficiently strong not only compared to $1/\tau$, but also the coherent dynamics \cite{Honer2011,ParisMandoki2017}.
The dephasing is dominated by atomic motion, with additional contributions from elastic scattering of the Rydberg electron by ground state atoms \cite{Baur2014,Gaj2014,Mirgorodskiy2017} and the AC-Stark shift induced by the trapping light.
To characterise the system, we determine $\kappa, \gamma_D$, and $\Gamma$ by comparing the observed transmission to the predictions of a model of $n_{\mathrm{sub}}$ effective three-level atoms strongly coupled to a chiral waveguide (Methods), assuming that $\kappa$, $\gamma_D$, and $\Gamma$ are equal for all absorbers.
The results of the model are in good agreement with the experiment for both the subtracted photons (Fig. 2a) and pulse shape of the transmitted light (Fig. 2b) for $\{\kappa,\Gamma,\gamma_D\}=\{0.494,0.045,2.329\}\,\mathrm{\mu s}^{-1}$ for $n_{\mathrm{sub}}=1$, $\{0.330,0.020,3.215\}\,\mathrm{\mu s}^{-1}$ for $n_{\mathrm{sub}}=2$, and $\{0.350,0.040,2.393\}\,\mathrm{\mu s}^{-1}$ for $n_{\mathrm{sub}}=3$.

A closer look at the number of subtracted photons (inset in Fig. 2a) reveals that it slightly exceeds $n_{\mathrm{sub}}$ at high $n_{\mathrm{in}}$ indicating that a single absorber may subtract multiple photons.
This excess cannot be explained by the deexcitation of the absorbers in the Rabi oscillation cycle as the associated re-emission occurs back into the probe mode in forward direction with rate $\kappa$.
Instead, it can be attributed to the small, but non-zero Raman decay $0<\Gamma<\kappa,\gamma_D$, which leads to spontaneous re-emission in a random, rather than the forward direction.
Its presence, even if weak, however, leads to reduced fidelity to subtract \textit{exactly} one photon per absorber as we will discuss in detail in our theoretical analysis.

To further demonstrate manipulation of the quantum state of light at the single photon level, we also investigate the effect of the subtraction on the correlations between transmitted photons.
Figure 2c shows the second order correlation function $g^{(2)}(t_1,t_2)=\langle n(t_1)n(t_2)\rangle/(\langle n(t_1)\rangle\langle n(t_2)\rangle)$ for two of the transmitted pulses shown in Fig. 2b and $n_{\mathrm{sub}}=3$ alongside the theory prediction for three cascaded three-level absorbers.
We observe $g^{(2)}(|t_2-t_1|=0) > 1$, i.e. bunching of the transmitted light. This bunching is expected as the subtraction operation reduces the mean photon number by $n_{\mathrm{sub}}=3$, without reducing the width of the photon number distribution compared to the incoming coherent pulses, thus inducing super-Poissonian statistics \cite{ParisMandoki2017,Stiesdal2018}.
It also highlights that subtraction of exactly $n_{\mathrm{sub}}$ photons is not to be confused with $n_{\mathrm{sub}}$ consecutive applications of annihilation operators, which also reduces $\langle n_{\mathrm{in}}\rangle$ of the incoming coherent pulses by up to $n_{\mathrm{sub}}$, but maintains Poisson statistics, or probabilistic subtraction \cite{Barnett2018}.
The correlations disappear towards the end of the pulses due to two effects: 
As saturation suppresses photon absorption, the incoming coherent light is transmitted without change and in addition absorbers are more likely to have undergone random Raman decay, which introduces Poissonian fluctuations in the superatom state that ultimately become reflected in the photon statistics.
Except for the rise- and fall-time of the pulse, where the experimental signal is dominated by noise, we observe good agreement between theory and experiment.

\begin{figure}
    \centering
    \includegraphics[width=\columnwidth]{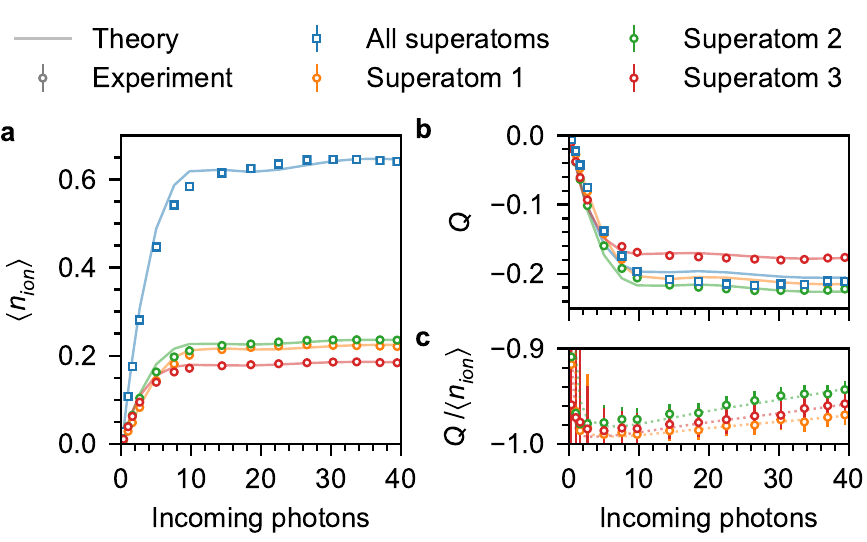}
    \caption{\textbf{Probing the absorber states via field ionisation of atoms in $\lvert r\rangle$.} The blue squares correspond to the total detected ions, while the yellow, green, and red circles correspond to the statistics for the first, second, and third absorber only. \textbf{a} Mean detected ions $\langle n_{\mathrm{ion}}\rangle$ vs. mean incoming photons for $n_{\mathrm{sub}}=3$. The variation in the detected ions for the individual absorbers are the result of a position-dependent detection efficiency (Methods). The shaded line shows the Rydberg population predicted by the three-level model normalised by the corresponding detection efficiency. \textbf{b} Mandel-Q parameter $Q$ for the same ion data as in a. \textbf{c} Ratio of $Q/\langle n_{\mathrm{ion}}\rangle$ for the individual absorbers. The dotted line shows the results of the model with added noise due to double excitation and dark counts.}
    \label{fig:figure3}
\end{figure}

To complement the transmission measurements, we also detect whether the absorbers are in the ground or a collective Rydberg state after the probe pulse by field ionisation of atoms in $\lvert r\rangle$. 
By resolving the time-of-flight from the atomic clouds to detection on a multi-channel plate (Fig. 1a), we can determine from which absorbers the produced ions originate.
Figure 3a shows mean detected ions $\langle n_{\mathrm{ion}}\rangle$ per pulse and absorber for $n_{\mathrm{sub}}=3$. The number of detection events from each absorber saturates at the detection efficiency $\eta$ as expected if no more than one excitation is supported, while the total number of ions saturates at the sum of the three absorbers.
The slight deviations in $\langle n_{\mathrm{ion}}\rangle$ for the individual absorbers result from a slight dependence of $\eta$ on the superatom position (between $0.18$ and $0.25$, Methods).
We also compare $\langle n_{\mathrm{ion}}\rangle$ to the combined populations of $\lvert W\rangle$ and $\lvert D\rangle$ after the probe pulse in the three-level model, again in good agreement with the experimental data following multiplication with the correspending values of $\eta$ .
To verify that each absorber is saturated by exactly one excitation, we analyse the counting statistics via the Mandel-$Q$ parameter $Q=\operatorname{Var}{n_{\mathrm{ion}}}/\langle n_{\mathrm{ion}}\rangle-1$ (Fig. 3b), which gives $-\eta$ for perfect blockade, $0$ for Poissonian, and $>0$ for super-Poissonian statistics.
Analysing each absorber separately, we find $Q\approx-\eta$ for sufficient number of input photons, as expected for saturation at one, while analysis of the combined counts from all absorbers yields $Q\approx-\langle n_{\mathrm{ion}}\rangle/n_{\mathrm{sub}}$ as expected for saturation at $n_{\mathrm{sub}}$ excitations.

Finally, Fig. 3c shows the ratio $Q/\langle n_{\mathrm{ion}}\rangle$ for the three individual absorbers. For large input photon number $\langle n_{\mathrm{in}} \rangle$ and perfect blockade, when each absorber contains exactly one excitation, this quantity should give $-1$. We observe a small deviation as $\langle n_{\mathrm{in}} \rangle$ increases which indicates the possibility of additional Rydberg excitations beyond the number of absorbers.
We account for these in the theoretical analysis by increasing the Rydberg populations obtained in the the three-level model with a small, photon-number dependent probability $p_2\langle n_{\mathrm{in}}\rangle$, which is independent of the superatoms' states.
In addition, we also account for the influence of dark counts in the ion detection by adding an offset independent of $\langle n_{\mathrm{in}}\rangle$, based on the experimentally observed dark-count rate of $9\,\mathrm{kHz}$.
These fluctuations induce a small Poissonian component in the ion counting statistics and thus shift $Q/\langle n_{\mathrm{ion}}\rangle$ to values above $-1$.
The results based on the modified values for $\langle n_{\mathrm{ion}}\rangle$ and $\operatorname{Var}{n_{\mathrm{ion}}}$ are shown in Fig. 3c and reproduce the experimental results well for $p_2 = 3.5, 6.5$ and $5.0\times 10^{-4}$ for the first, second, and third absorber respectively. To achieve good agreement for the second absorber, we need to increase the constant noise by a factor $5$ compared to the dark count rate.
This is presumably due to the occasional detection of ions originating from the first and third absorber as well as atoms trapped in between the superatoms in the corresponding time-of-flight window attributed to the second absorber, which is expected to be more prone to these events due to its central positon.

Our analysis underpins the hypothesis that there is a small, $\langle n_{\mathrm{in}}\rangle$-dependent probability to create additional Rydberg excitations, which may be caused by several mechanisms which we cannot distinguish in our experiment due to the small magnitude of the effect. First, residual atoms trapped between the superatom ensembles may be excited to $\vert r\rangle$ if $\langle n_{\mathrm{in}}\rangle$ is sufficiently high that powerbroadening becomes comparable to the AC-Stark shift induced by the tightly confining optical traps. Second, imperfections in the blockade can occur from interaction-induced pair-state resonances on shells within the blockaded volume \cite{Derevianko2015}, similar to the anti-blockade effect \cite{Ates2007,Amthor2010}. 

\noindent\textbf{Parameter optimisation and scalability.}
In the following theoretical analysis, we determine the optimal parameters for the superatom photon absorber and discuss the potential for scaling  beyond $n_{\mathrm{sub}}=3$.
We base the discussion on the results of the Lindblad master equation for a one-dimensional chain of chirally coupled superatoms described by the three-level model~\cite{Pichler2015,Stiesdal2018,Kramer2018} (Methods).
On one hand, parameter optimisation is necessary as the Raman decay $\Gamma$ introduces a probabilistic component into the otherwise deterministic scheme and sets an upper limit on the pulse length $\tau$.
For short $\tau$ on the other hand, $\gamma_D$ must be balanced with the driving strength $\sqrt{\kappa R_\mathrm{in}}$ to yield high absorption probability.

\begin{figure*}
    \centering
    \includegraphics[width=\textwidth]{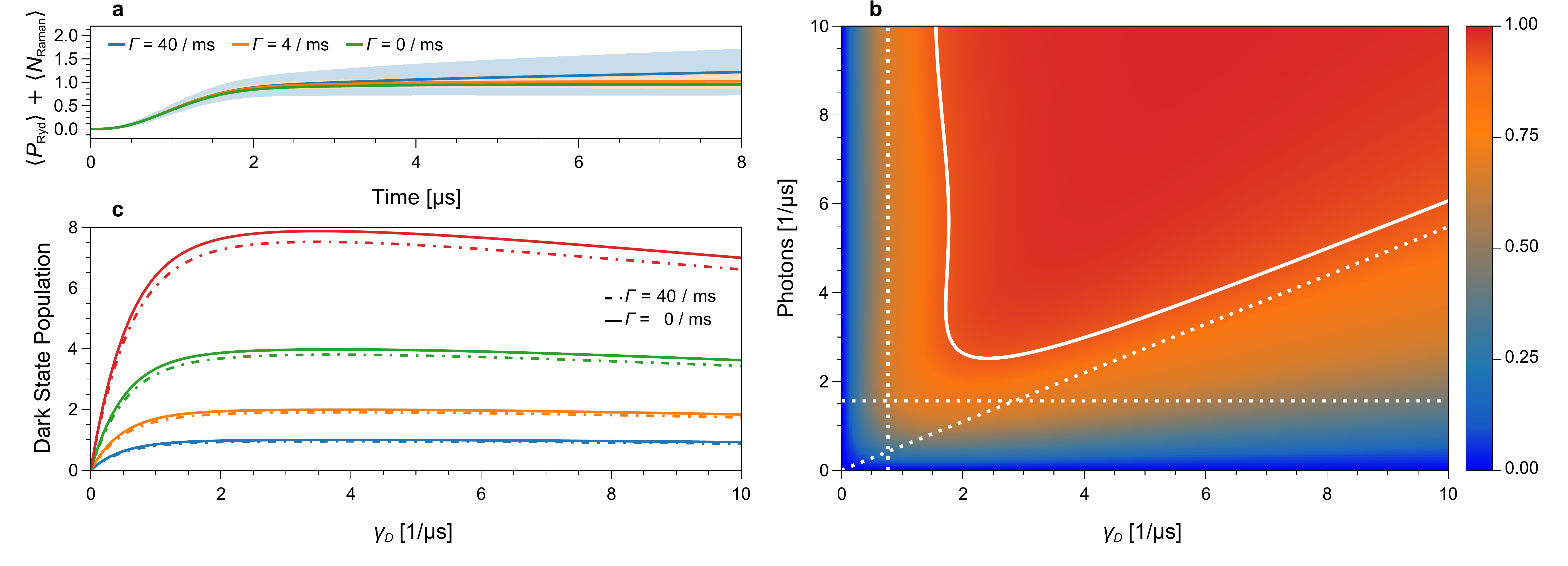}
    \caption{\textbf{Theory prediction for the superatom dynamics}.
        All figures are at $\kappa = \SI{0,35}{/ \mu s}$.
        \textbf{a} Number of subtracted and Raman emitted photons for a constant pulse $R_\mathrm{in} = \SI{5}{Photons / \mu s}$ and with dephasing rate $\gamma_D = \SI{2,4}{/ \mu s}$.
        Shaded regions indicate one standard deviation of the number of Raman emitted photons.
        \textbf{b} Dark state population after driving a single superatom for $\tau = \SI{3}{\mu s}$ without Raman decay $\Gamma = 0$.
        The solid line indicates the 90\,\% level.
        The dashed horizontal, vertical and diagonal lines correspond to $\sqrt{\kappa R_\mathrm{in}} \tau = \pi / 2$, $\exp(-\gamma_D \tau) = 0.1$ and $\exp(-4\kappa R_\mathrm{in}\tau / \gamma_D) = 0.1$, respectively.
        \textbf{c)} Mean dark state population in a chain of 1, 2, 4 and 8 superatoms after driving the superatoms for $\tau = \SI{4}{\mu s}$ at a constant rate of $R_\mathrm{in} = \SI{5}{Photons / \mu s}$.
    }
    \label{fig:figure4}
\end{figure*}

We begin our analysis by considering stochastic loss of photons due to incoherent Raman decay from the Rydberg manifold into $\vert G\rangle$.
It opens a scattering channel into non-observed modes, with a mean number of lost photons
\begin{equation}
    \langle N_\mathrm{Raman}(t)\rangle = \Gamma \int_{t_0}^t dt'\, P_\mathrm{Ryd}(t'),
\end{equation}
where $P_\mathrm{Ryd}(t)$ is the combined population of $\vert W\rangle$ and $\vert D\rangle$.
Figure~\ref{fig:figure4}a shows the total number of absorbed photons $P_\mathrm{Ryd}(t) + \langle N_\mathrm{Raman}(t)\rangle$ together with one standard deviation $\sqrt{\mathrm{var}(N_\mathrm{Raman}(t))}$ (Methods) for $\Gamma = 0$, $0.004$, and $0.04\,\mathrm{\mu s^{-1}}$, for a single superatom at constant driving. 
The fluctuations in photon absorption increase over time, highlighting that deterministic photon subtraction requires both low $\Gamma$ and short durations $\tau$.

Consequently, we now analyse the dynamics of a single superatom at $\tau = 3\,\mathrm{\mu s}$, similar to the experiment.
Figure~\ref{fig:figure4}b shows the population of $\vert D\rangle$ vs. $\gamma_D$ and $R_{\mathrm{in}}$ indicating a large parameter regime where photon absorption occurs with high probability.
This regime is bounded by three processes with independent time scales, which we indicate by dashed lines.
Firstly, the superatom is excited into $\vert W\rangle$ with rate $\sqrt{\kappa R_\mathrm{in}}$ and the requirement $\sqrt{\kappa R_\mathrm{in}}\tau \gg 1$ gives a lower bound for the necessary photon rate.
Similarly, $\vert W\rangle$ decays into $\vert D\rangle$ with rate $\gamma_D$ and thus the dark state will only be populated for $\gamma_D \tau \gg 1$.
However, at large $\gamma_D$, the superatom dynamics enter an overdamped regime in which we can adiabatically eliminate $\vert W\rangle$ (Methods) and the effective absorption rate scales asymptotically as $\gamma_\mathrm{eff} \simeq 4\kappa R_\mathrm{in} / \gamma_D$.
Therefore we also require $\gamma_\mathrm{eff} \tau \gg 1$, limiting the maximal dephasing rate $\gamma_D$.

Lastly, we solve the Master equation for a chain of $n_{\mathrm{sub}}$ driven superatoms and show that the number of subtracted photons scales well with $n_{\mathrm{sub}}$.
Figure~\ref{fig:figure4}c compares the total dark state population for an ideal system with $\Gamma = 0$ (solid lines) to $\Gamma = 0.04\,\mathrm{\mu s^{-1}}$ (dashed lines) and shows the potential to extend our photon subtraction scheme up to $n_{\mathrm{sub}} = 8$.
In the simulation, we drive the superatoms with a mean number of $20$ photons, which indicates that our setup can work well even when the number of photons becomes comparable to $n_{\mathrm{sub}}$.
The dark state population never reaches $n_{\mathrm{sub}}$ exactly, which is due to the short pulse duration~$\tau$ considered here and, in the absence of Raman decay, the absorption probability can be made arbitrarily large by increasing $\tau$.

%=================%
%  Conclusion
%=================%

\section*{Discussion}

The theoretical analysis of our experimental results reveals that the main contributions to imperfections in the subtracted photon number are two-fold and are not necessarily unique to our scheme.
First, a finite lifetime of the saturated state leads to excess absorption and the random nature of a decay process like Raman decay in our system introduces a probabilistic component into an initially deterministic scheme.
This applies to any scheme which employs excited or metastable states and the severity of the impact depends on the decay strength compared to the absorber-photon coupling and the pulse duration.
While the loss through Raman decay may initially seem as a disadvantage of our implementation, it should be noted that insertion loss into waveguides and cavities can lead to similar probabilistic fluctuations.   
Second, the slight deviation from $n_{\mathrm{sub}}$ for coherent input pulses of finite duration is more general and affects all subtraction schemes relying on irreversible transfer into a dark state or separate optical modes irrespective the absorber nature, also including hybrid systems of waveguide- and cavity-coupled single quantum emitters \cite{Rosenblum2016,Du2020}.

In our scheme, Raman decay could be further suppressed by reducing either $\Omega_c$ or increasing $\Delta$, with the latter also reducing residual absorption on the probe transition.
To compensate the associated reduction in $\kappa$, one can increase the number of atoms $N$ per superatom via the ensemble density or increase the probe waist with $r_{\mathrm{B}}$ as upper constraint.
Meanwhile, for increasing $\kappa$ combined with fine-tuning of $\gamma_D$, the dark-state population converges towards $n_{\mathrm{sub}}$ shifting the curves in Fig. 4c upwards.
In this context, performance limitations will ultimately occur for high $R_{\mathrm{in}}$ as powerbroadening causes a breakdown of the blockade.

While high-fidelity preparation of quantum states of light may require more substantial performance improvements, limitations are less stringent for other applications of our setup.
An immediate application is number-resolved detection of up to $n_{\mathrm{sub}}$ photons based on the number of absorbers in a Rydberg state.
Currently, performance would be limited by the low efficiency $\eta$ to detect the superatom state via field ionisation, but this could be significantly improved by replacing the MCP by another model or using optical detection \cite{Olmos2011,Gunter2012,Gunter2013}.
By increasing $n_{\mathrm{sub}}$ well beyond the expected photon number, a weak photon-absorber coupling $\kappa$ could also be compensated.
Meanwhile, Raman decay should still be strongly suppressed as it reduces the detection efficiency for each absorber.

In summary, we have experimentally demonstrated a first implementation of controlled multi-photon subtraction from weak coherent probe pulses using a cascaded chain of saturable Rydberg superatom absorbers in free space.
Our theoretical analysis has identified both technical and fundamental sources of imperfections, including the introduction of probabilistic fluctuations into an in principle deterministic scheme through decay and a probability slightly below unity to transfer an absorber into its saturated state for coherent input pulses of finite duration.  

Obvious next steps include improving the subtraction fidelity via the measures discussed above, changes to the optical trapping of the atomic ensembles to further increase $n_{\mathrm{sub}}$, and the implementation of optical readout for number-resolved photon detection.
More generally, our system of cascaded superatoms is also well suited to study the behaviour of emitters coupled to a chiral waveguide \cite{Mahmoodian2018,Prasad2020,Pichler2015}, as the superatoms not only introduce photon correlations, but the photons also coherently mediate interactions between the superatoms, which should become evident in the limit $\gamma_D\ll\kappa$ and with increasing $n_{\mathrm{sub}}$.

\section*{Methods}

\noindent\textbf{Ensemble preparation.} 
We start from a cigar shaped ensemble of $^{87}$Rb atoms in a crossed optical dipole trap (wavelength $1070\,\mathrm{nm}$, $1/e^2$-waist $\approx 55\,\mathrm{\mu m}$, intersection angle $30^{\circ}$) loaded from a magneto-optical trap (MOT).
Following a final compression of the MOT, the atoms are evaporatively cooled as we reduce the trap light intensity in two stages.
For additional cooling and to reduce atom loss, we employ Raman sideband cooling for $16\,\mathrm{ms}$ during each of the linear evaporation ramps.
To create multiple ensembles for multiple superatom absorbers, we generate multiple tightly focused optical traps with an elliptical cross-section (wavelength $805\,\mathrm{nm}$, $1/e^2$-waists $\approx 9\,\mathrm{\mu m}$ along and $\approx 29\,\mathrm{\mu m}$ perpendicular to the probe)  that intersect perpendicularly with the cigar-shaped ensemble as well as the probe and control beams by feeding several radio-frequency signals into an AOD (as shown in Fig. 1 and discussed in the main text). 
An objective system, translates the resulting differences in diffraction angle into different trap positions that can be tuned via the signals' frequencies over a range of order $100\,\mathrm{\mu m}$, which is limited by the axial extend of the crossed region of the reservoir trap.
In our experiments the separation between the ensemble centres is $\approx 75\,\mathrm{\mu m}$ for $n_{\mathrm{sub}}=2$ and $\approx 50\,\mathrm{\mu m}$ for $n_{\mathrm{sub}}=3$.

Before experiments, we ramp the crossed dipole trap intensity to zero to release atoms confined between the dimples before increasing it again to provide confinement in the radial probe direction for the superatom ensembles.
In combination with the $1/e^2$-waist radius of the probe ($\approx 6.5\,\mathrm{\mu m}$), the dimple confinement restricts the excitation volume below the blockade range.
The focus of the control beam is larger ($\approx 14\,\mathrm{\mu m}$) to limit variation of $\Omega_c$ across the excitation volume.\\

\noindent\textbf{Experimental sequence.} 
Following the preparation outlined above, we turn the crossed dipole trap off for $14\,\mathrm{\mu s}$ every $100\,\mathrm{\mu s}$, while the dimple traps are left on to maintain confinement along the probe direction.
The resulting AC-Stark shift is compensated by adjusting the probe frequency accordingly and we ensure that all superatom absorbers have the same resonance frequency by individual fine-tuning via the power for each RF signal applied to the AOD.
Besides the axial confinement, the AC-Stark shift also helps to suppress Rydberg excitation of residual atoms trapped in between the dimple potentials.
Following each single experimental shot, we field-ionise any Rydberg atoms to gain information  about the absorber state and avoid the presence of residual Rydberg excitations during the next iteration of the superatom excitation. 
The ions are detected on a multi-channel plate (MCP).
In total, we repeat the cycle described above $500$ times before releasing the atoms to obtain reference pulses of the probe light in the same manner in absence of any atoms and preparing new atomic ensembles.\\

\noindent\textbf{Site-resolved ion detection.}
In order to attribute the ions detected following each experimental shot to Rydberg excitations in different ensembles, we use a time of flight method.
We find that the pulses generated by the MCP detector following detection of an ion occur in a time window with a width of $\approx 30\,\mathrm{ns}$ and a typical separation of several $10\,\mathrm{ns}$ between the arrival times for two superatom ensembles separated by $50\,\mathrm{\mu m}$. 
Combined with the $3\,\mathrm{ns}$ time resolution of our data acquisition, this allows us to attribute a detection windows of $75\,\mathrm{ns}$ to the location of each individual superatoms. 
In Fig. 3, the detection efficiency varies between $\eta \approx 0.18$ and $0.25$ depending on the position of a superatom and is generally highest in the single absorber case. 
The variation is caused by a grid of steel wires, which is placed in front of the detector to shield the atoms in the experimental region from the strong electric field produced by the MCP front plate and partially obstructs the ion trajectories.
For a single superatom, the applied ionisation and steering fields can be adapted to minimise the influence of the grid, but for multiple absorbers we cannot avoid that a fraction of ions is blocked by the wires, which depends on the location of their origin.\\

\noindent\textbf{Theoretical Description.}

We describe each superatom $i$ as an effective 3-level atom whose ground state $|G_i\rangle$ is coupled to a collective excited state, the bright state, $|W_i\rangle$ by the coherent probe field $\alpha(t)$ (with $R_\mathrm{in} = |\alpha(t)|^2$).
The photon absorber relies on shelving excitations into a non-radiating dark state $|D_i\rangle$, which we model by an incoherent decay of the bright state with rate $\gamma_D$. Assuming the dipole and rotating wave approximation and no Raman decay, the superatoms obey the master equation~\cite{Pichler2015}
\begin{align}
    \notag
    \partial_t \rho &= -\frac{i}{\hbar}[H_\mathrm{drive}(t) + H_\mathrm{exc}, \rho]
    + \kappa\mathcal{D}\left[\sum_{i=1}^N \sigma_{W_i}^-\right]\rho \\
    % \notag
    &\qquad + \gamma_D \sum_{i=1}^N \mathcal{D}[\sigma_{D_i}^+\sigma_{W_i}^-]\rho.
    % &\qquad + \Gamma \sum_{i=1}^N \Big( \mathcal{D}[\sigma_{W_i}^-] \rho +
    % \mathcal{D}[\sigma_{D_i}^-]\rho\Big).
    \label{eq:MasterEquation}
\end{align}
The Master equation consists of the action of the probe field
\begin{equation}
    H_\mathrm{drive} = \sqrt{\kappa} \sum_{i=1}^N
    \Big(\alpha(t) \sigma_{W_i}^+ + \alpha^*(t) \sigma_{W_i}^- \Big),
\end{equation}
a hopping term due to the exchange of virtual photons
\begin{equation}
    H_\mathrm{exc} = -\frac{i\kappa}{2} \sum_{i>j}
    \Big( \sigma_{W_i}^+ \sigma_{W_j}^- - \sigma_{W_j}^+ \sigma_{W_i}^- \Big)
\end{equation}
and the dissipative decay terms $\mathcal{D}\left[\sum_{i=1}^N \sigma_{W_i}^-\right]$ and $\sum_{i=1}^N \mathcal{D}[\sigma_{D_i}^+\sigma_{W_i}^-]$, describing the collective decay of the superatoms and depashing of each bright state into the respective dark state.
We use the notation $\sigma_{A_i}^- \equiv \vert G_i\rangle\langle A_i\vert$, $\sigma_{A_i}^+ \equiv \vert A_i\rangle\langle G_i\vert$ and $\mathcal{D}[\sigma] \rho \equiv \sigma \rho \sigma^\dagger - \{ \sigma^\dagger \sigma, \rho \}/2$.

To understand non-deterministic effects in our photon absorber setup it is essential to include Raman decay in our model.
In the simplest description, Raman decay enters as an additional decay term $\Gamma( \mathcal{D}[\sigma_{W_i}^-]\rho + \mathcal{D}[\sigma_{D_i}^-]\rho)$ for each superatom.
To gain access to the number statistics of the emitted Raman photons, we further introduce a virtual spin chain to our model and modify the Raman decay terms so that each Raman decay yields an excitation in the spin chain. 
This allows us to calculate the standard deviation of the number of emitted Raman photons, as depicted as in Figure~\ref{fig:figure4}a.

At large $\gamma_D$, the superatoms enter an overdamped regime and further increasing $\gamma_D$ negatively impacts the photon absorption rate.
In the overdamped dynamics, we may adiabatically eliminate the bright state from the Master equation, which we achieve by setting $\partial_t \rho_{WW} = 0 = \partial_t \rho_{WG} = (\partial_t \rho_{GW})^*$, with the short hand notation $\langle A\vert\rho\vert B \rangle = \rho_{AB}$.
Under the adiabatic elimination, the Master equation for a single superatom reduces to a classical rate equation for the ground state and dark state population
\begin{align}
    \partial_t \rho_{GG} &=          -  \gamma_\mathrm{eff} \rho_{GG}, \\
    \partial_t \rho_{DD} &= \phantom{-} \gamma_\mathrm{eff} \rho_{GG},
\end{align}
where the effective decay rate reads as
\begin{equation}
    \gamma_\mathrm{eff} =
    \frac{4\kappa R_\mathrm{in} \gamma_D}{(\kappa + \Gamma + \gamma_D)^2 + 4 \kappa R_\mathrm{in}}.
    \label{eq:EffectiveDecay}
\end{equation}

\section*{Data availability}
The data presented in this paper are available at [DOI to be added in proof].

\bibliographystyle{naturemag}
\bibliography{references}

\section*{Acknowledgements}
This work has received funding from the European Union's Horizon 2020 program under the ERC consolidator grants SIRPOL (grant no. 681208) and RYD-QNLO (grant no. 771417), the ErBeStA project (grant no. 800942), and under grant agreement no. 845218 (Marie Sk\l{}odowska-Curie Individual Fellowship to H.B.), the Deutsche Forschungsgemeinschaft (DFG) under SPP 1929 GiRyd project BU 2247/4-1, and the Carlsberg Foundation through the Semper Ardens Research Project QCooL.

\section*{Author contributions}
N. S., H. B., M. G. H., and S. H. performed the experimental work, K. K., J. K., and H. P. B. conducted the theoretical analysis. All authors contributed to  data analysis, discussion of the results, and preparation of the manuscript.

\section*{Competing interests}
The authors declare no competing interests.

\end{document}